\documentclass[10pt]{iopart}

\usepackage{iopams}
\usepackage[utf8]{inputenc}
\usepackage{color}
\usepackage{graphicx}
\usepackage{overpic}
\usepackage{siunitx}

\usepackage{cite}
\usepackage{hyperref}
\hypersetup{
	colorlinks=true,
	pdfstartview=FitV,
	linkcolor=blue,
	citecolor=red,
	urlcolor=blue,
	pdfauthor={HOPPE Mathias, mhop@kth.se},
	pdfsubject={Benign termination upper pressure limit},
	pdftitle={Benign termination upper pressure limit}
}

\newcommand{\dd}{\mathrm{d}}

\newcommand{\eqref}[1]{(\ref{#1})}

\newcommand{\pnB}{p_{\rm n}^{\rm B}}

\newcommand{\vA}{v_{\rm A}}

\begin{document}
	\title[Benign termination upper pressure limit]
	{An upper pressure limit for low-Z benign termination of runaway electron beams in TCV}

		\author{
		M.~Hoppe$^1$,
		J.~Decker$^2$, U.~Sheikh$^2$, S.~Coda$^2$, C.~Colandrea$^2$,
		B.~Duval$^2$, O.~Ficker$^3$, P.~Halldestam$^4$, S.~Jachmich$^5$,
		M.~Lehnen$^{5,\ddagger}$, H.~Reimerdes$^2$, C.~Paz-Soldan$^6$, M.~Pedrini$^2$,
		C.~Reux$^7$, L.~Simons$^2$, B.~Vincent$^2$, T.~Wijkamp$^{8,9}$,
		M.~Zurita$^2$, the TCV team$^\dagger$ and the EUROfusion Tokamak
		Exploitation Team$^\star$
	}

	\address{%
		$^1$Department of Electrical Engineering, Royal Institute of Technology, SE-11428 Stockholm, Sweden\\
		$^2$Swiss Plasma Center, Ecole Polytechnique Fédérale de Lausanne, CH-1015 Lausanne, Switzerland\\
		$^3$Institute of Plasma Physics of the CAS, Za Slovankou 3, 182 00 Prague 8, Czech Republic\\
		$^4$Max Planck Institute for Plasma Physics, D-85748 Garching, Germany\\
		$^5$ITER Organization, Route de Vinon-sur-Verdon, CS 90 046, 13067 St. Paul Lez Durance Cedex, France\\
		$^6$Department of Applied Physics and Applied Mathematics, Columbia University, New York, New York 10027, USA\\
		$^7$CEA-IRFM, F-13108 Saint-Paul-les-Durance, France\\
		$^8$Department of Applied Physics, Eindhoven University of Technology, Eindhoven 5600 MB, Netherlands\\
		$^9$DIFFER---Dutch Institute for Fundamental Energy Research, De Zaale 20, 5612 AJ Eindhoven, Netherlands\\
		$^\dagger$see the author list of B.P.~Duval {\em et al} 2024 Nucl.~Fusion 64 112023\\
		$^\star$see the author list of E.~Joffrin {\em et al} 2024 Nucl.~Fusion 64 112019\\
		$^\ddagger$deceased
	}
	\ead{mhop@kth.se}

	\vspace{10pt}
	\begin{indented}
		\item[]November 2024
	\end{indented}

	\begin{abstract}
		We present a model for the particle balance in the post-disruption
		runaway electron plateau phase of a tokamak discharge. The model
		is constructed with the help of, and applied to, experimental data from
		TCV discharges investigating the so-called ``low-Z benign termination''
		runaway electron mitigation scheme. In the benign termination scheme,
		the free electron density is first reduced in order for a subsequently
		induced MHD instability to grow rapidly and spread the runaway electrons
		widely across the wall. We show that the observed non-monotonic
		dependence of the free electron density with the measured neutral
		pressure is due to plasma re-ionization induced by runaway electron
		impact ionization. At higher neutral pressures, more target particles
		are present in the plasma for runaway electrons to collide with and
		ionize. Parameter scans are conducted to clarify the role of the runaway
		electron density and energy on the free electron density, and it is found
		that only the runaway electron density has a noticeable impact. While
		the free electron density is shown to be related to the spread of heat
		fluxes at termination, the exact cause for the upper neutral pressure
		limit remains undetermined and an object for further study.
	\end{abstract}

	\submitto{\PPCF}
	\ioptwocol

	\section{Introduction}\label{sec:intro}
	Plasma disruptions are one of the most pressing issues for reactor-scale
	tokamaks~\cite{Hender2007,Lehnen2015}. Tokamak disruptions are
	associated with a range of threats to the plasma facing components,
	including strong heat fluxes to the wall, large currents and forces induced
	in toroidally closed structures, and beams of runaway electrons which
	can cause severe, localized heat damage. While mitigation strategies relying
	on massive material injection have been developed over past decades, recent
	simulations suggest that it may be difficult to mitigate all types of damage
	in a disruption simultaneously using such
	techniques~\cite{Pusztai2023,Ekmark2024}. Furthermore, simulations suggest
	that in the nuclear phase of ITER operation, large runaway electron currents
	may be unavoidable should a disruption
	occur~\cite{Vallhagen2022,Vallhagen2024}.

	A complementary technique for runaway electron mitigation, commonly referred
	to as ``benign termination'', has recently received much
	attention~\cite{Reux2021,PazSoldan2021}. The term refers to the termination
	of a full beam of runaway electrons, which is considered benign if the
	runaways are spread widely across a large wall area, rather than depositing
	their energy in a localized spot. To achieve this, the benign termination
	scheme aims to trigger a large-scale, rapidly growing MHD instability after
	the primary disruption has occurred and the plasma is dominated by a beam of
	runaway electrons~\cite{Bandaru2021}. In practice, this is achieved by
	injecting low-$Z$ material, after the disruption, to reduce the plasma
	temperature and density (through recombination) until the Alfvén speed
	$v_{\rm A}$ is large enough for fast MHD mode growth~\cite{Hollmann2023}.
	The fast mode growth ensures that all runaway electrons are expelled before
	avalanche multiplication can reconstitute the runaway electron
	current~\cite{McDevitt2023a}. Once the plasma is largely recombined, the
	safety factor is reduced---often as a consequence of natural or induced
	plasma compression at constant plasma current---whereupon a large-scale MHD
	instability is triggered that terminates the runaway electron beam.

	Several devices have contributed to the effort of developing the
	benign termination
	scheme~\cite{Reux2021,PazSoldan2021,Carnevale2021,Sheikh2024}. In
	DIII-D, JET and TCV---the three devices on which the most extensive
	investigations have been carried out---it is found that, in addition to the
	lower limit on the amount of low-$Z$ impurities required to achieve
	recombination, there is also an upper limit above which the MHD mode growth 
	is reduced and runaway electron beam termination is no longer benign. In the
	present paper, we will utilise this extensive dataset for runaway electron
	beam benign termination obtained with TCV to analyse the upper limit in
	low-$Z$ material injected amount observed experimentally.

	We begin this paper by describing the TCV scenario used to acquire the
	dataset in section~\ref{sec:experiment}. In section~\ref{sec:theory},
	details of the model are then presented, and in section~\ref{sec:results} it
	is shown to account for the observed density variations with neutral
	pressure of TCV plasmas. Alongside the TCV results, the sensitivity to
	runaway electron density and energy is also investigated. Finally,
	section~\ref{sec:conclusions} summarizes our main conclusions and discusses
	implications for future devices.

	\section{Experimental setup}\label{sec:experiment}
	The experiments reported in this paper have been conducted on the
	\emph{Tokamak à Configuration Variable} (TCV), located at the Swiss Plasma
	Center in Lausanne, Switzerland~\cite{Duval2024}. The standard benign
	termination scenario on TCV starts from a quiescent flat-top plasma with a
	runaway electron seed population, as described in \cite{Decker2022}.
	Once the runaway electron seed has been established, around $t=\SI{0.7}{s}$,
	a disruption is deliberately triggered via a neon (Ne) or argon (Ar) massive
	gas injection (MGI). The MGI causes the plasma temperature to drop to around
	$\sim\SI{5}{eV}$ and the plasma current to be overtaken entirely by runaway
	electrons. After the disruption, additional deuterium (D) or hydrogen (H)
	particles are injected using MGI and the electron density declines
	sharply, indicating that sufficiently low temperatures for recombination
	have been achieved. The partially recombined plasma is then held steady to
	ensure that reliable measurements of plasma parameters can be obtained, and
	fuelling valves are used to compensate for wall pumping in order to maintain
	the neutral pressure until the plasma is compressed against the inner
	wall. The compression is done while maintaining a constant plasma current
	and until $q_{\rm edge}=2$ is reached, at which point an MHD instability is
	triggered and terminates the runaway electron beam.

						Some of the key diagnostics used in studying benign termination on TCV
	include the Thomson scattering system for measuring electron temperature and
	density, the baratron gauges for measuring neutral pressure, and IR cameras
	for estimating the runaway electron heat fluxes to the wall. A few of the
	channels of the Thomson scattering system are equipped with filters suitable
	for measuring electron temperatures in the single digit eV range, and in
	this work we use data from one such channel observing near the centre of
	the plasma ($r\approx\SI{10}{cm}$). The measured signal is averaged over the
	time window after MGI and fuelling during which the plasma parameters are
	held steady.

	The baratron gauges are situated outside the toroidal field coils and
	are connected to the vacuum vessel via dedicated extension tubes. Neutral
	particles must therefore make their way through the tube before the neutral
	pressure can be recorded. As such, the neutral pressure measured and
	quoted in this analysis is significantly different from the neutral
	pressure in the centre of the plasma. However, the relevance of the neutral
	pressure as an observable lies in its relation to the neutral particle
	content close to the wall, which therefore characterises the plasma state
	prior to runaway electron beam termination. This was experimentally
	demonstrated in \cite{Sheikh2024}.

				To estimate the impact of a beam termination, the surface area of
	the runaway electron wall heating is estimated from camera images captured
	with an IRCAM Equus 81k M infrared (IR) camera with a radial view to the
	plasma. The procedure for estimating the spread of the heat was described in
	detail in \cite{Sheikh2024} and relies on counting the number of pixels
	which see ``significant'' heating after the termination, yielding a
	\emph{wetted area} measure which will be used here to quantify whether a
	given runaway electron beam termination is benign or not.

		\begin{figure}
		\begin{overpic}[width=\columnwidth]{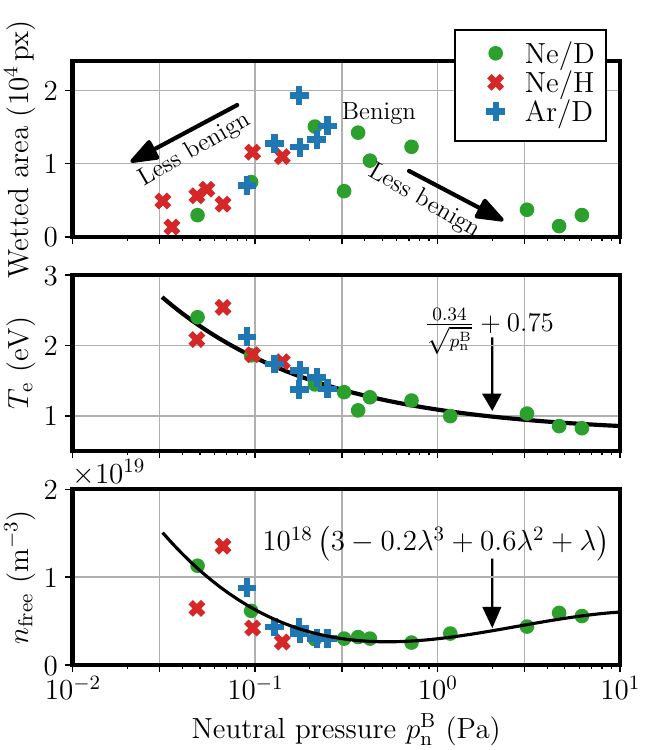}
			\put(11,88){(a)}
			\put(11,59){(b)}
			\put(11,31){(c)}
		\end{overpic}
		\caption{%
			Variation of
			(a) wetted area,
			(b) electron temperature and
			(c) free electron density,
			as functions of neutral pressure. Temperature and density are
			measured using Thomson scattering. Runaway electron beam termination
			is considered benign if the wetted area is large, which is achieved
			at intermediate neutral pressures. Each data point corresponds to
			one TCV discharge. Temperatures and density values are averaged in
			time after the disruption, prior to the plasma compression.
		}
		\label{fig:experiment}
	\end{figure}

	\subsection{Experimental results}
\begin{figure}
		\centering
		\includegraphics[width=\columnwidth]{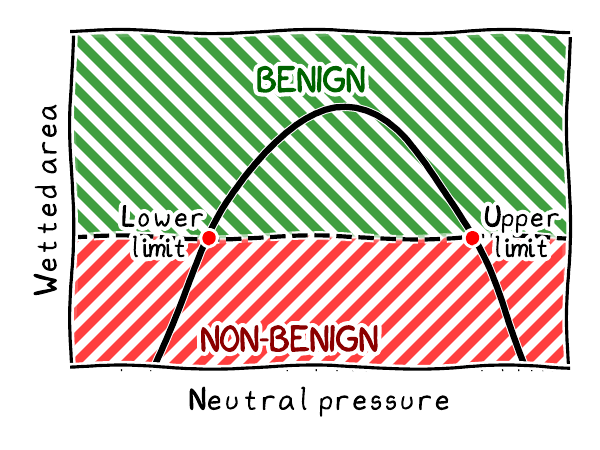}
		\caption{
			Conceptual sketch of the upper and lower neutral pressure limits.
			The benign termination scheme assumes that the wetted area must
			exceed a certain value in order to avoid damage to the wall. Our
			experimental results show that benign termination is then possible
			within a finite range of pressures.
		}
		\label{fig:concept}
	\end{figure}

	In \cite{Sheikh2024}, analysis of a set of benign termination
	experiments conducted on TCV was presented. The experiments were aimed at
	exploring the physics of benign termination at different neutral pressures,
	and with different hydrogenic and impurity species. Data for this same set
	of experiments is shown in figure~\ref{fig:experiment}. Panel (a) shows the
	wetted area, estimated from IR camera images, immediately after runaway
	electron beam termination, which indicates that maximum spreading occurs for
	a measured neutral pressure of $\pnB\approx\SI{0.3}{Pa}$ (``most benign
	scenario''). Assuming that the wetted area must be greater than some value
	$A_{\rm crit}$ to avoid damage to the plasma-facing components, the
	non-monotonic nature of the wetted area indicates that two limits---a lower
	and an upper limit---exist for the neutral pressure, as illustrated in
	figure~\ref{fig:concept}. For a termination to be considered benign, the
	neutral pressure must take values between these two limits. Panels (b) and
	(c) show the central electron temperature and density, time-averaged over
	the phase during which the plasma is held steady before the final runaway
	electron beam termination.

	A plausible explanation for the low (high) wetted area is the slow (fast)
	MHD mode growth resulting from a small (large) Alfvén speed
	$\vA=B/\sqrt{\mu_0\rho}$, where $B$ is the magnetic field strength, $\mu_0$
	the permeability of free space, and $\rho$ is the plasma mass
	density~\cite{PazSoldan2021,Sheikh2024}. While $\rho$ for a weakly ionized
	plasma has contributions from both ionized and neutral
	particles~\cite{Vranjes2008}, we find that the electron density alone well
	follows the trend of the wetted area. Hence, low electron densities are
	associated with large wetted areas and vice versa. The non-monotonic
	variation of the electron density is explained as follows: at low neutral
	pressures, the plasma temperature remains high so that the plasma ionization
	by thermal electron impact is similarly high. As the target neutral pressure
	is raised, corresponding to an increase in the injected number of hydrogenic
	particles, the temperature gradually decreases, causing the plasma to
	gradually recombine and thereby reducing $\rho$. However, as the neutral
	pressure is increased beyond $\pnB\approx\SI{0.3}{Pa}$, the plasma density
	rises, whereas the temperature continues to decrease. As we will show in the
	present paper, this phenomenon can be explained by the ionization induced by
	runaway electrons as they collide with neutrals and partially ionized atoms.

	\section{Model}\label{sec:theory}
\begin{figure}
		\centering
		\includegraphics[width=\columnwidth]{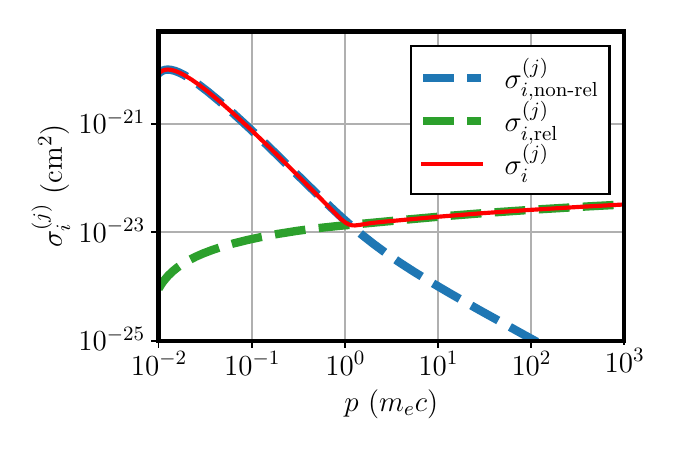}
		\caption{
			Fast electron impact ionization cross-section for collisions with
			neutral D/H, as function of the fast electron total momentum (in
			normalized units).
		}
		\label{fig:xsection}
	\end{figure}

	To describe the observations at higher neutral pressures, we develop a
	particle balance model which accounts for the ionization caused by runaway
	electrons. In a collisional-radiative model, the time rate of change of the
	density $n_i^{(j)}$ of ion species $i$ in charge state $j$ is expressed as
	\begin{eqnarray}
		\frac{\dd n_i^{(j)}}{\dd t} &=
			\left[
				I_i^{(j-1)} n_e + \mathcal{I}_i^{(j-1)}
			\right] n_i^{(j-1)}
			+
			R_i^{(j+1)}n_en_i^{(j+1)}
			-\nonumber\\
			&-
			\left[
				\left( I_i^{(j)} + R_i^{(j)} \right) n_e
				+ \mathcal{I}_i^{(j)}
			\right] n_i^{(j)},
			\label{eq:pb}
	\end{eqnarray}
	where, $n_e$ is the free electron density, $I_i^{(j)}$ and $R_i^{(j)}$
	denote the ionization and recombination rate coefficients, respectively, and
	$\mathcal{I}_i^{(j)}$ is the rate of fast electron impact ionization. In
	this work, $I_i^{(j)}$ and $R_i^{(j)}$ are taken from
	OPEN-ADAS~\cite{ADAS,OPENADAS}, while the fast electron impact ionization
	rate is calculated from
	\begin{equation}\label{eq:Ire}
		\mathcal{I}_i^{(j)} =
			\int\dd p\,
				p^2 v\sigma_i^{(j)}(p) f_{\rm re}(p).
	\end{equation}
	Here, $p$ denotes the electron momentum normalized to the electron mass
	and speed-of-light $m_ec$, $v = cp/\sqrt{1+p^2}$ is the electron speed, and
	$f_{\rm re}$ is the electron momentum distribution function (integrated over
	the pitch and gyro angles). The fast electron impact ionization
	cross-section $\sigma_i^{(j)}$ for collisions with species $i$ in charge
	state $j$ is given by~\cite{Garland2020}
	\begin{equation}\label{eq:sigma}
		\sigma_i^{(j)}(p) =
			\left[1-s(p)\right]\sigma_{i,\text{non-rel}}^{(j)}(p) +
			s(p)\sigma_{i,\rm rel}^{(j)}(p),
	\end{equation}
	where $s(p)$ is an interpolation coefficient (modified compared to
	Ref.~\cite{Garland2020} for better agreement in the non-relativistic limit)
								\begin{equation}
		s(p) = \left(
			1 + \frac{5}{W_{\rm kin}(p)}\exp\left[-5.11W_{\rm kin}(p)\right]
		\right)^{-1},
	\end{equation}
	with $W_{\rm kin}=p^2/(1+\sqrt{p^2+1})$ the runaway electron kinetic energy
	in normalized units. The cross-section $\sigma_{i,\text{non-rel}}^{(j)}$ is
	the Burgess-Chidichimo model, valid at non-relativistic energies, and
	$\sigma_{i,\rm rel}^{(j)}$ is a relativistic correction. These components
	of the cross-section are given by
	\begin{numparts}
	\begin{eqnarray}
		\sigma_{i,\text{non-rel}}^{(j)}(p) =
			\pi C a_0^2
			\left(\frac{\mathrm{Ry}}{\Delta W_{\rm iz}}\right)^2
			\frac{1}{U}\left(\log U\right)^{1+\beta^\star/U},\label{eq:sigmaNR}\\
		\sigma_{i,\rm rel}^{(j)}(p) =
			\pi C a_0^2\alpha^2
			\frac{\mathrm{Ry}}{\Delta W_{\rm iz}}\left(
				\log{\frac{m_ec^2p^2}{2\Delta W_{\rm iz}}} -
				\frac{p^2}{p^2+1}
			\right).\label{eq:sigmaR}
	\end{eqnarray}
	\end{numparts}
	In these expressions, $a_0$ denotes the Bohr radius, $\alpha\approx 1/137$
	the fine-structure constant, $\mathrm{Ry}$ is the Rydberg energy, and
	\begin{equation}
				U = \frac{m_ec^2}{\Delta W_{\rm iz}} W_{\rm kin}.
	\end{equation}
	The factor $C$, near-threshold modification factor $\beta^\star$, and
	effective ionization potential $\Delta W_{\rm iz}$ in
	equations~\eqref{eq:sigmaNR} and~\eqref{eq:sigmaR} all depend on the
	particular ion species and charge state being considered. In this work we
	follow the approach of \cite{Hoppe2021} and treat $C$, $\beta^\star$
	and $\Delta W_{\rm iz}$ as free parameters, fitting them to the OPEN-ADAS
	data such that the ionization rates $I_i^{(j)}$ are recovered when
	$f_{\rm re}$ in equation~\eqref{eq:Ire} is substituted for a Maxwellian.
	Figure~\ref{fig:xsection} shows the energy dependence of the full fast
	electron impact ionization cross-section and its components.

\begin{figure}
		\centering
		\includegraphics[width=\columnwidth]{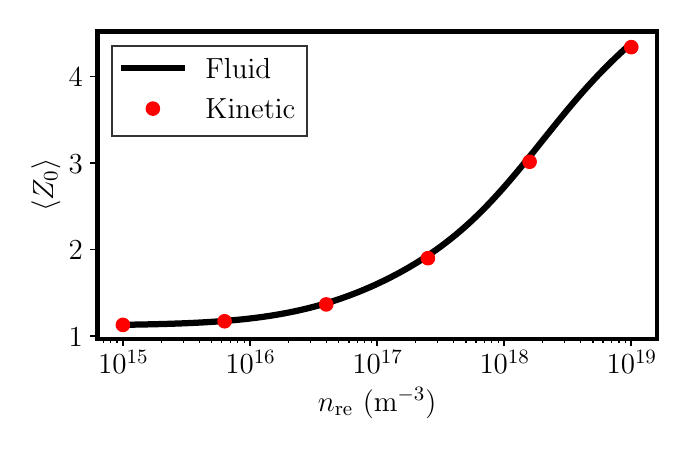}
		\caption{
			Comparison of the effective charge in a pure Ar plasma at
			$T_e=\SI{1}{eV}$ in the presence of runaway electrons with density
			$n_{\rm re}$. The solid line shows the result with the fluid runaway
			electron ionization model~\eqref{eq:fluidIre}, while the red dots
			are the result of a simulation where $f_{\rm re}$ in
			equation~\eqref{eq:Ire} is taken as the analytical avalanche
			distribution of equation~(2.17) in Ref.~\cite{Embreus2018}.
		}
		\label{fig:ICScomparison}
	\end{figure}

	Evaluation of the fast electron impact ionization rate $\mathcal{I}_i^{(j)}$
	also requires information about the fast electron momentum distribution
	function, $f_{\rm re}(p)$. Due to the logarithmic energy dependence of
	$\sigma_{i,\rm rel}^{(j)}$, it is however not always necessary to accurately
	know the fast electron energy distribution. Instead, we may take
	$f_{\rm re}$ to be a delta function at a characteristic fast electron
	momentum $p_{\rm re}$ and evaluate the ionization rate for it. We take
	\begin{equation}
		f_{\rm re}(p) = \frac{n_{\rm re}}{p^2}\delta\left(p-p_{\rm re}\right),
	\end{equation}
	where $n_{\rm re}$ is the number density of fast electrons. Substituting
	this into equation~\eqref{eq:Ire}, and taking the relativistic limit
	($p_{\rm re}\gg 1$) of equation~\eqref{eq:sigma}, we finally arrive at
	\begin{equation}\label{eq:fluidIre}
		\mathcal{I}_i^{(j)} =
			\pi C a_0^2\alpha^2 c n_{\rm re}
			\frac{\mathrm{Ry}}{\Delta W_{\rm iz}}\left[
				\log\left(\frac{m_ec^2p_{\rm re}^2}{2\Delta W_{\rm iz}}\right) - 1
			\right].
	\end{equation}
	The logarithmic dependence on $p_{\rm re}$ means that its exact value is 
	unimportant, enabling this expression for the runaway electron impact
	ionization to be used in, for example, fluid models. For the remainder of
	this paper we will use a representative value of
	$p_{\rm re}=20m_ec\approx\SI{10}{MeV}/c$. Figure~\ref{fig:ICScomparison}
	shows that equation~\eqref{eq:fluidIre} agrees well with
	equation~\eqref{eq:Ire} evaluated with the full
	cross-section~\eqref{eq:sigma} and the analytical runaway electron
	distribution function of equation (2.17) in \cite{Embreus2018}.

	In this work, we will consider the particle balance in a steady-state
	post-disruption plasma and assume $\dd n_i^{(j)}/\dd t\to 0$ in
	equation~\eqref{eq:pb}. When taking this limit, one additional constraint
	per ion species must be imposed to close the system of equations. For the
	impurity species (Ne or Ar), we impose the number density $n_{\rm Ne/Ar,0}$
	of Ne or Ar particles in the plasma:
	\begin{equation}
		\sum_j n_{\rm Ne/Ar}^{(j)} = n_{\rm Ne/Ar,0}.
	\end{equation}
	For the main ion species (H or D) we impose a constraint on the free
	electron density from experimental measurements
	\begin{equation}\label{eq:quasineutral}
		\sum_i\sum_{j=1}^{Z_i} jn_i^{(j)} = n_e.
	\end{equation}
						
	\section{Results}\label{sec:results}
								In this section we apply the model described in section~\ref{sec:theory} to
	analyse TCV data. Using empirical fits of electron temperature and density,
	we show that runaway electron impact is the dominant ionization mechanism
	at higher neutral pressures, and discuss the consequences. Finally, we study
	the sensitivity of the free electron density to the runaway electron
	parameters.

	\subsection{Temperature, density, and runaway density}\label{sec:fits}
	The measured electron temperature and density, as functions of neutral
	pressure in the baratron $\pnB$, are shown in figure~\ref{fig:experiment}.
	Given the small vertical spread in the data, curves can be well fitted
	to both quantities. For the electron temperature
	\begin{equation}\label{eq:fit_Te}
		T_e\left(\pnB\right) = \frac{0.34}{\sqrt{\pnB}} + 0.75,
	\end{equation}
	with $\pnB$ given in units of Pa. For the electron density
	\begin{equation}\label{eq:fit_ne}
		n_e\left(\pnB\right) = 10^{18}\,\si{\per\meter\cubed}\left(
			3 - 0.2\lambda^3 + 0.6\lambda^2 + \lambda
		\right),
	\end{equation}
	where $\lambda\equiv\log\pnB$.

	The runaway electron density can be estimated from the plasma current. In
	the TCV scenarios considered here, the plasma current was
	$I_{\rm p} = \SI{150}{kA}$ and since the temperature is very low in the
	plateau phase of a disruption, no appreciable ohmic current can be driven,
	implying that all of this current is carried by runaway electrons. Assuming
	a spatially uniform density with runaway electrons travelling at
	relativistic speeds and zero pitch angle, the runaway electron density can
	be estimated from
	\begin{equation}
		I_{\rm p} = ecn_{\rm re} A = ecn_{\rm re} \pi a^2,
	\end{equation}
	where $e$ is the elementary charge and $a$ denotes the plasma minor radius.
	With the condition that the edge safety factor $q_{\rm edge}=2$ (which is
	satisfied right before the termination and translates to an upper limit on
	$n_{\rm re}$), the plasma minor radius is
	$a^2 = \mu_0R_0I_{\rm p}/(\pi B_0)$~\cite{Wesson}, where $R_0=\SI{88}{cm}$
	is the plasma major radius and $B_0=\SI{1.45}{T}$ the on-axis magnetic
	field strength. This yields a runaway electron density of
	$n_{\rm re}=\SI{2.7e16}{\per\meter\cubed}$, which will be used as the
	baseline value in this section.

	In the scenarios with Ne, a total of $7.2\times 10^{18}$ particles are
	injected into the vacuum vessel with volume $4.632\,\si{m^3}$. In Ar
	scenarios, a slightly higher $7.5\times 10^{18}$ particles are injected.
	In the calculations of this paper, we assume that the Ne/Ar density is
	spread uniformly throughout the vessel, and that 10\% of the injected
	particles end up in the plasma centre.

	\subsection{Ionization rate}
	Only for the lowest neutral pressure, where $T_e$ is relatively high, are
	collisions between thermal particles the dominant ionization effect. For
	$\pnB > \SI{0.05}{Pa}$, runaway electron impact becomes the primary source
	of free electrons.

\begin{figure}
		\centering
		\includegraphics[width=\columnwidth]{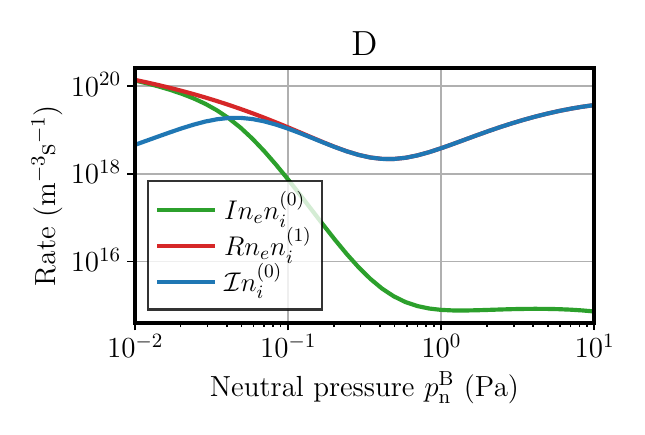}
		\caption{
			Ionization rate $I n_e n_i^{(0)}$, recombination rate
			$R n_e n_i^{(1)}$, and fast electron impact ionization rate
			$\mathcal{I} n_i^{(0)}$ for deuterium, as a function of
			the neutral pressure $\pnB$ as measured in the baratron.
		}
		\label{fig:rates}
	\end{figure}

	With the empirical fits~\eqref{eq:fit_Te} and~\eqref{eq:fit_ne}, using the
	latter as a constraint on the particle balance equation~\eqref{eq:pb}, we
	can evaluate the ionization and recombination rates as functions of neutral
	pressure, as shown for D in figure~\ref{fig:rates}. Having established that
	runaway electron impact is the dominant ionization mechanism, we can infer
	that the rise in free electron density observed in the experimental data in
	figure~\ref{fig:experiment} is a result of the fact that the number of
	collisions between runaway electrons and D/H particles increases with
	neutral pressure.

	The local minimum in the ionization rates around $\pnB=\SI{0.4}{Pa}$ in
	figure~\ref{fig:rates} coincides approximately with the point at which Ne
	fully recombines, in the absence of runaway electrons. Up until this point,
	the neutral pressure in the baratron can increase as a result of Ne
	recombining. Beyond $\pnB=\SI{0.4}{Pa}$, all Ne outside of the plasma is
	recombined and the only way for the neutral pressure to further increase is
	by raising the D or H particle content in the vessel. More particles in the
	vessel means more targets for the runaway electrons to collide with and
	ionize, and as a consequence, the free electron density will continue to
	increase with neutral pressure.

\begin{figure}
		\centering
		\includegraphics[width=\columnwidth]{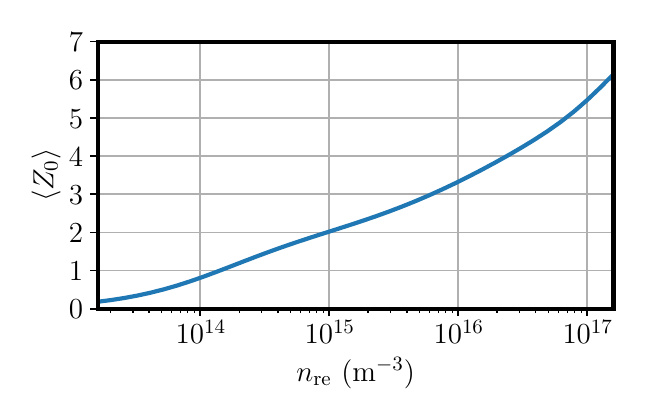}
		\caption{
			Average charge state of Ne as a function of the runaway electron
			density, at a temperature $T_e=\SI{1.23}{eV}$. In the absence of
			any runaway electrons, Ne would be fully recombined at this
			temperature, but due to the strong runaway electron ionization
			the particles remain partially ionized.
		}
		\label{fig:avgZ0}
	\end{figure}

		The model indicates that the runaway electron impact ionization has a
	major impact on the ionization degree of the Ne inside the plasma.
	Figure~\ref{fig:avgZ0} shows the average charge state of Ne as a function of
	the runaway electron density in a $T_e=\SI{1.23}{eV}$ plasma. Without
	runaway electrons, Ne would be fully recombined. Accounting for the runaway
	electron ionization term~\eqref{eq:Ire}, the average charge may reach almost
	$\langle Z_0\rangle=4$, a value which otherwise requires $T_e=\SI{18}{eV}$.
	This implies that measurements of $\langle Z_0\rangle$ for Ne, or other
	impurity atoms, could be used to validate the fast-electron impact
	ionization cross-section~\eqref{eq:sigma} in a tokamak, by measuring the
	relative abundance of particles in different charge states.
				
	\subsection{Sensitivity to runaway parameters}\label{sec:reparams}
	With the help of the model in section~\ref{sec:theory}, the sensitivity of
	the free electron density $n_e$ to the runaway electron density $n_{\rm re}$
	and momentum $p_{\rm re}$ can be analysed. We begin by considering the
	effect of runaway density on the ionization rates $I_i^{(j)}n_en_i^{(0)}$
	and $\mathcal{I}_i^{(j)}n_i^{(0)}$.

\begin{figure}
		\centering
		\includegraphics[width=\columnwidth]{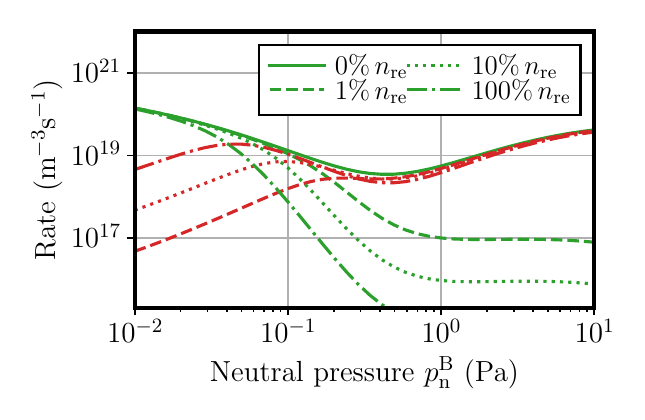}
		\caption{
			Effect of the runaway electron density $n_{\rm re}$ on the thermal
			$I_i^{(j)}n_en_i^{(0)}$ (green) and runaway electron
			$\mathcal{I}_i^{(j)}n_i^{(0)}$ (red) ionization rates. The runaway
			density is given as a percentage of the baseline value
			$n_{\rm re}=\SI{2.7e16}{\per\meter\cubed}$.
		}
		\label{fig:nrescan_rates}
	\end{figure}

	In figure~\ref{fig:nrescan_rates}, these are shown for TCV as functions of
	the measured neutral pressure, again employing the empirical
	fits~\eqref{eq:fit_Te} and~\eqref{eq:fit_ne} to relate the rates to neutral
	pressure. Since the rates are calculated assuming a prescribed electron
	density, $n_e = n_e(\pnB)$, a reduction in $n_{\rm re}$ and the runaway
	electron ionization rate means that more D/H must be added to match the
	observed $n_e$. The thermal ionization rate $I_i^{(j)}$ consequently
	increases as $n_{\rm re}$ is reduced, and makes up an increasingly larger
	fraction of the total ionization. Nevertheless, even at $n_{\rm re}=1\%$ of
	the baseline value the runaway electron ionization dominates, and thermal
	ionization does not contribute substantially to the ionization of the
	plasma. Only when the runaway electron ionization effect is removed from the
	model can the thermal ionization provide a significant number of---and in
	this case all of the---free electrons in the plasma, and only after a large
	amount of D/H particles have been inserted into the simulation. We recall
	that the amount of D/H in the plasma is a free parameter in our calculations
	since it cannot be constrained from the available experimental data.

\begin{figure}
		\centering
		\includegraphics[width=\columnwidth]{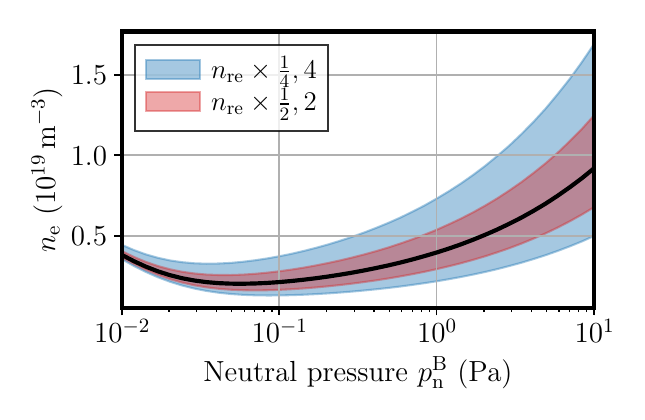}
		\caption{
			Variation of the free electron density as the runaway electron
			density $n_{\rm re}$ is changed by factors of 2 (1/2) and 4 (1/4).
			At $\pnB=\SI{1}{Pa}$, a doubling (halving) of $n_{\rm re}$ results
			in a $29\%$ higher ($21\%$ lower) free electron density, while a
			fourfold increase (reduction) results in a $67\%$ increase
			($37\%$ decrease) in $n_e$.
		}
		\label{fig:nrescan_nfree}
	\end{figure}

	The effect of $n_{\rm re}$ on the free electron density $n_e$ is shown in
	figure~\ref{fig:nrescan_nfree}, where $n_{\rm re}$ is varied by factors of
	2 and 4. As expected from figure~\ref{fig:nrescan_rates}, $n_{\rm re}$
	has a negligible effect on $n_e$ at low neutral pressures. Towards higher
	pressure the effect becomes more pronounced, and at a pressure of
	$\SI{1}{Pa}$ a $\times 2$ ($\times 1/2$) variation in $n_{\rm re}$ yields a
	$+36\%$ ($-26\%$) variation in $n_e$. The fact that the runaway electron
	impact ionization rates in figure~\ref{fig:nrescan_rates} converge at
	higher pressures is a consequence of the constraint~\eqref{eq:quasineutral}
	on the free electron density which we impose. Since this constraint adjusts
	the D/H content in the simulation such that the experimentally measured
	electron density is always recovered, the total ionization rate must always
	yield the experimental electron density, which at high pressures means that
	the runaway electron ionization source term must have the same value
	regardless of the runaway electron density.

\begin{figure}
		\centering
		\includegraphics[width=\columnwidth]{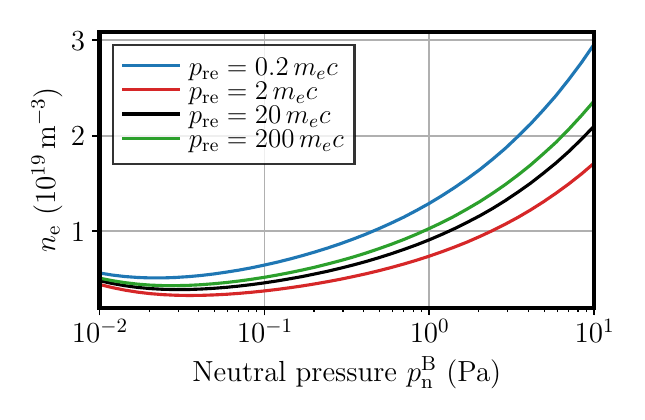}
		\caption{
			Variation of the free electron density as the runaway electron
			total momentum $p_{\rm re}$ is changed by factors of 10. At a
			pressure of $\pnB=\SI{1}{Pa}$, as the momentum is increased from 
			$2m_ec$ to $20m_ec$, the free electron density increases by $17\%$,
			while the rise in density from $p_{\rm re}=20m_ec$ to
			$p_{\rm re}=200m_ec$ is $12\%$.
		}
		\label{fig:prescan_nfree}
	\end{figure}

	As discussed in section~\ref{sec:theory}, the energy of the runaway
	electrons has a very small effect on the ionization source strength
	$\mathcal{I}$. In figure~\ref{fig:prescan_nfree}, the impact on the free
	electron density is shown as the runaway energy is varied over three orders
	of magnitude. At a pressure of $\pnB=\SI{1}{Pa}$, a ten-fold increase in
	momentum from $p=2m_ec$ to $p=20m_ec$ yields only a $19\%$ increase in
	$n_e$, while a further increase to $p=200m_ec$ yields another $13\%$ rise in
	$n_e$. One point at $p=0.2m_ec$ illustrates how at very low runaway momenta,
	the ionization rate increases with decreasing runaway density due to the
	enhanced collisionality of those runaways. Order unity variations in runaway
	electron energy gives very small variations in the free electron density,
	suggesting that this effect can be neglected in most practical situations.

	Altogether, this suggests that the runaway electron density is the most
	crucial parameter for the plasma ionization at high pressures. In terms of
	directly measurable parameters, the runaway electron density is most closely
	related to the plasma current (and, specifically, the current density), and
	so the plasma ionization in the plateau will depend on the plateau current.

	\section{Discussion and conclusions}\label{sec:conclusions}
	In this paper we have presented a model for the particle balance in a
	post-disruption tokamak plasma and applied it to a benign termination
	scenario in the TCV tokamak. We find that runaway electron impact ionization
	is by far the dominant ionization mechanism at play, and that it gives rise
	to the increase in electron density at high neutral pressures reported in
	several experimental benign termination
	studies~\cite{PazSoldan2021,Sheikh2024}. Since runaway electron ionization
	is independent of the plasma temperature, and increases with the D/H
	particle density, higher neutral pressures result in more ionizing
	collisions and higher electron densities. While this density rise is shown
	to be associated with an upper neutral pressure limit for the benign
	termination, the exact cause of the pressure limit cannot be concluded
	from the analysis conducted here.

	The runaway electron impact ionization source term primarily depends on the
	runaway density, and the density and type of the target particle. The source
	is linear in both densities, which means that both parameters have a
	relatively strong, albeit non-linear, impact on the resulting free electron
	density. Conversely, the ionization source is found to depend only
	weakly on the runaway electron energy. The ionization cross-section, and
	consequently the ionization source term, depends logarithmically on the
	energy, requiring order-of-magnitude differences in runaway energy for any
	substantial difference in the source term. This motivates a mono-energetic
	approximation for the runaway electron distribution and enables the
	inclusion of the ionization source term in fluid models such
	as~\cite{Hollmann2019,Bandaru2021}, where the runaway electron energy
	distribution is not evolved.

		Previous simulations indicate that molecules play a major role for the
	plasma evolution in post-disruption plateaus~\cite{Hollmann2023}. In this
	work, we have for simplicity neglected the formation of molecules and their
	interaction with runaway electrons, however this should not affect our main
	conclusions. As shown in~\cite{Hollmann2023}, ionized molecules tend to
	recombine at a faster rate than ionized atoms. Molecules could therefore
	contribute to enhancing the recombination rate compared to what is used
	here, thereby changing the predicted free electron density at a given
	neutral pressure $\pnB$. This will however not change the picture that
	runaway electron impact ionization is the dominant ionization mechanism, but
	will rather only affect our estimated D/H content in the plasma.

	It remains unclear where the upper neutral pressure limit will be located
	in reactor-scale tokamaks, as the model presented here does not account for
	all effects which come into play in the termination of a runaway electron
	beam. Crucially, to determine the upper limit, a model for the electron
	temperature would be needed~\cite{McDevitt2023b}, as well as a model for
	the MHD mode growth rate. Fluid codes could address at least some of these
	questions, and the ionization source term~\eqref{eq:fluidIre} is a
	convenient component to include in such codes in order to capture the
	apparently dominant runaway electron impact ionization physics.

	Although the model presented here does not allow any firm conclusions to be
	drawn about the situation in reactor-scale tokamaks, it does give some
	insights into how the runaway electron impact ionization source might be
	affected. In particular, by considering the runaway electron density and
	plasma size at the time of termination, with $q_{\rm edge}=2$, the same
	argument as used in section~\ref{sec:fits} gives that
	$n_{\rm re}\propto j_{\rm re}\propto B_0/R_0$, where $B_0$ is the on-axis
	magnetic field strength and $R_0$ the plasma major radius. For TCV, this
	ratio is $B_{0,\rm TCV}/R_{0,\rm TCV}\approx\SI{1.65}{T/m}$ while for ITER
	it is $B_{0,\rm ITER}/R_{0,\rm ITER}\approx\SI{0.85}{T/m}$. This suggests
	that the ionization source term may be roughly half as strong in ITER, for
	the same hydrogen content, compared to TCV. On the other hand, the relation
	between particle content and neutral pressure in ITER is unknown, as is the
	relation between electron density and the MHD mode growth rate required for
	benign termination in ITER. Thus, large uncertainties remain before firm
	conclusions about the efficacy of benign termination in ITER can be drawn.

	\section*{Acknowledgements}
		This work has been carried out within the framework of the EUROfusion
	Consortium, partially funded by the European Union via the Euratom Research
	and Training Programme (Grant Agreement No 101052200 -- EUROfusion). The
	Swiss contribution to this work has been funded by the Swiss State
	Secretariat for Education, Research and Innovation (SERI). Views and
	opinions expressed are however those of the author(s) only and do not
	necessarily reflect those of the European Union, the European Commission or
	SERI. Neither the European Union nor the European Commission nor SERI can be
	held responsible for them. This work was supported in part by the Swiss
	National Science Foundation. C.P-S.\ acknowledges support from the US
	Department of Energy under award DE-SC0022270.

	\bibliography{ref}

	\appendix
	
	\section{Ionization cross-section fitting parameters}
	The ionization cross-sections~\eqref{eq:sigmaNR} and~\eqref{eq:sigmaR}
	depend on the parameters $C$, $\Delta W_{\rm iz}$, and $\beta^\star$. In
	this paper, we treat them as free parameters and choose them such that the
	ionization source term~\eqref{eq:Ire} recovers the ADAS ionization rate
	when equation~\eqref{eq:Ire} is evaluated with a Maxwellian distribution
	function. Doing so for H/D, Ne, and Ar yields the parameter values listed
	in table~\ref{tab:parameters}. Coefficients for additional species, and
	tools for doing the fits to ADAS data, are distributed with the DREAM
	code~\cite{Hoppe2021}.

	\begin{table}[ht!]
		\centering
		\caption{
			Values determined for the free parameters of the fast electron
			impact ionization cross section $\sigma_i^{(j)}$, for the different
			charge states $j=Z_0$ of D/H, Ne, and Ar.
		}
		\label{tab:parameters}
		\begin{tabular}{lc|c|c|c}
			\textbf{Species} & $Z_0$ & $C$ & $\Delta W_{\rm iz}$ (eV) & $\beta^\star$\\\hline
			\textbf{D/H} & 0 & $3.053$ & $13.60$ & $0.2988$\\\hline
			\textbf{Ne} & 0 & $4.592$ & $21.56$ & $0.7987$\\
			& 1 & $26.11$ & $40.96$ & $0.4521$\\
			& 2 & $19.69$ & $63.42$ & $0.7721$\\
			& 3 & $20.83$ & $97.19$ & $0.0000$\\
			& 4 & $15.69$ & $126.2$ & $0.0000$\\
			& 5 & $5.037$ & $157.9$ & $0.0810$\\
			& 6 & $7.264$ & $207.3$ & $0.0591$\\
			& 7 & $3.905$ & $239.1$ & $0.0000$\\
			& 8 & $8.188$ & $1196$ & $0.0000$\\
			& 9 & $5.139$ & $1362$ & $0.0000$\\\hline
			\textbf{Ar} & 0 & $18.88$ & $15.76$ & $0.6036$\\
			& 1 & $19.91$ & $27.63$ & $0.2788$\\
			& 2 & $18.57$ & $40.73$ & $0.2068$\\
			& 3 & $16.28$ & $59.58$ & $0.1942$\\
			& 4 & $12.77$ & $74.84$ & $0.2334$\\
			& 5 & $9.701$ & $91.29$ & $0.3559$\\
			& 6 & $11.07$ & $124.4$ & $0.1992$\\
			& 7 & $7.508$ & $143.5$ & $0.4796$\\
			& 8 & $28.84$ & $422.6$ & $0.0890$\\
			& 9 & $25.47$ & $479.8$ & $0.1088$\\
			& 10 & $21.79$ & $540.4$ & $0.1224$\\
			& 11 & $18.46$ & $619.0$ & $0.1565$\\
			& 12 & $14.80$ & $685.5$ & $0.2181$\\
			& 13 & $11.08$ & $755.1$ & $0.2634$\\
			& 14 & $8.305$ & $855.5$ & $0.0499$\\
			& 15 & $4.376$ & $918.4$ & $0.0842$\\
			& 16 & $7.996$ & $4120$ & $0.0256$\\
			& 17 & $4.000$ & $4426$ & $0.0246$\\\hline
		\end{tabular}
	\end{table}
\end{document}